\documentclass{appolb}
\usepackage{graphicx}
\usepackage{amsmath,amssymb,amsfonts,color,cite,soul}
\usepackage{subfigure}

\newcommand{\sd}{{\textsc{SecDec}}}

\newcommand{\rd}{{\mathrm{d}}}

\def\eps{\varepsilon}
\begin{document}
\title{Calculation of multi-scale, multi-loop integrals using SecDec 3%
\thanks{Presented at the XXXIX International Conference of Theoretical Physics ``Matter to the deepest'', September 13 - 18, 2015, Ustro\'{n}, Poland}%
}
\author{S.~Borowka
\address{Institute for Physics, University of Zurich \\Winterthurerstr.190,
8057 Zurich, Switzerland}
}
\maketitle
\begin{abstract}
In these proceedings 
the publicly available program \sd{} is briefly described. Its main virtues and new features are 
summarized, including suggestions for an optimal usage of the program. 
\end{abstract}
  
\section{Introduction}
In light of the increasing precision achieved by the experiments at the Large Hadron Collider, 
the uncertainties on theoretical predictions need to decrease at equal speed. 
Background processes need to be understood at the same level of accuracy as the 
predictions of signals so that the Standard Model can be confirmed or deviations from it be 
discovered. 
This requires the computation of higher orders in perturbation theory. 
For a considerable number of processes, predictions at NNLO would be desirable for LHC Run II and beyond.
These involve the computation 
of complicated, often massive, two-loop diagrams. 
In particular multi-loop multi-scale integrals are highly challenging for currently available analytical 
techniques. However, they are more easily accessible with numerical evaluation approaches. 
One public program to evaluate such diagrams is \sd{} 
\cite{Carter:2010hi,Borowka:2012yc,Borowka:2013cma,Borowka:2015mxa}, 
which relies on the method of sector decomposition \cite{Hepp:1966eg,Roth:1996pd,Binoth:2000ps} to 
factorize UV and IR singularities. Using subtraction terms, the singularities can be extracted and 
their coefficients integrated numerically. \sd{} is available from {\it http://secdec.hepforge.org/}. 
Other public implementations are the programs 
\textsc{sector\_decomposition} \cite{Bogner:2007cr,Gluza:2010rn} and 
\textsc{Fiesta} \cite{Smirnov:2008py,Smirnov:2009pb,Smirnov:2013eza,Smirnov:2015mct}. 
\section{Applicability of the program \sd}
The program \sd{} is designed for multifold applications. Many recent 
improvements were implemented for Feynman multi-loop integrals. Nonetheless, more 
general parametric integrals and integrals which deviate from the standard form of Feynman 
integrals can be computed as well. 

\medskip

A scalar Feynman integral $G$ in $D$ dimensions 
at $L$ loops with $N$ propagators, where 
the propagators can have arbitrary, not necessarily integer powers $\nu_j$,  
has the following representation in momentum space:
\begin{align}\label{eq:integraldef}
G=&\int\prod\limits_{l=1}^{L} \rd^D\kappa_l\;
\frac{1}
{\prod\limits_{j=1}^{N} P_{j}^{\nu_j}(\{k\},\{p\},m_j^2)}\;,\\
\rd^D\kappa_l=&(i\pi^{\frac{D}{2}})^{-1}\,\rd^D k_l\;,\;
P_j(\{k\},\{p\},m_j^2)=q_j^2-m_j^2+i\delta\;,
\end{align}
where the $q_j$ are linear combinations of external momenta $p_i$ and loop momenta $k_l$. 
Multi-loop Feynman integrals with scalar, rank $R$ or inverse propagator numerators 
can be handled by \sd{} within the {\it loop} setup. Prefactors dependent on the dimensional 
regulator $\eps$ are allowed, any scale dependence must be explicitly given by the user. 
Internally, the user input of the aforementioned loop integral specifications is 
transformed into a representation in terms of Feynman parameters. After loop momentum integration 
the general expression for a scalar Feynman integral reads \small
\begin{align}
\label{eq:scalar}
G = \frac{ (-1)^{N_{\nu}} }{ \prod_{j=1}^{N} \Gamma(\nu_j) } \Gamma( N_{\nu}-LD/2 )
\int\limits_{0}^{\infty} \!\prod \limits_{j=1}^{N} \text{d}x_j\,\,x_j^{\nu_j-1}\,\delta(1-\sum_{l=1}^N x_l) 
\frac{{\cal U}^{N_{\nu}-(L+1) D/2}(\vec{x})}{{\cal F}^{N_\nu-L D/2}(\vec{x},s_{ij})}\;\text{}
\end{align} \normalsize
with $N_\nu=\sum_{j=1}^N \nu_j$, see e.g. Ref.~\cite{Heinrich:2008si} for the definition 
of general Feynman loop integrals of rank $R$.
The graph polynomials {$\cal U$} and {$\cal F$} in Eq.~(\ref{eq:scalar}) contain the sub-UV and IR 
divergences of an integral, respectively. The polynomials can be constructed from topological cuts by \sd{} or  
from the explicit specification of propagators in momentum space in the {\it math.m} input file. 

\medskip

When numerical checks at intermediate stages of an analytical (multi-) loop calculation are 
of interest, or when the integral at hand is simply too complicated for a direct numerical 
evaluation, an analytical preparation can be beneficial.  Then, the integral is no longer 
of the form of Eq.~(\ref{eq:scalar}). Such integrals, as long as they match the following structure, \small
\begin{align}
\label{eq:userdefined}
G_{user} = P(\eps) \;
 \int\limits_{0}^{1} \prod \limits_{j=1}^{N} 
\text{d}x_j\,\,x_j^{a_j(\eps)} \;
{\cal N}(\vec{x}, s_{ij},\eps)\; 
{\cal U}^{ \text{expoU}(\eps )}(\vec{x}, s_{ij}) \;  
 {\cal F}^{ \text{expoF}(\eps )}(\vec{x}, s_{ij})\;\text{,}
\end{align} \normalsize
can be handled within the {\it userdefined} setup of \sd. In Eq.~(\ref{eq:userdefined}), the function 
${\cal N}$ may contain products of polynomials with either a direct or exponential dependence 
on $\eps$. The functions ${\cal U}$ and ${\cal F}$ may have negative exponents, and powers 
$a_j<0$ are allowed. The function ${\cal F}$ is chosen as reference for the analytical 
continuation of the integrand to the physical region. For further details, the reader is 
referred to Refs.~\cite{Borowka:2012yc,Borowka:2013cma,Borowka:2014aaa}. 

\medskip

More general parametric functions are handled in the {\it general} setup of 
\sd. Examples are phase space integrals, where IR divergences are 
regulated dimensionally, or hypergeometric functions. 
This setup allows for an arbitrary number of products in the integrand, where each of them 
can have a negative exponent. While the poles of the integrand are factorized using sector decomposition, 
it is sometimes of interest to include additional $\eps$-dependent functions, which may depend on 
the integration parameters but do not contain any non-factorized poles. 
These functions can be masked in so-called "dummy functions".
The latter feature is implemented 
in version 3 of the program. 
\section{Brief summary of the operational sequence}
The program \sd{} processes the user input further by factorizing the 
poles of the integrand using either one of the two heuristic \cite{Binoth:2000ps,Heinrich:2008si} 
iterated sector decomposition strategies or one of the two deterministic 
\cite{Kaneko:2009qx,Borowka:2015mxa} ones. The user can choose among the four, see 
Ref.~\cite{Borowka:2015mxa} for details. 
While the heuristic algorithms used to lead to more compact expressions  to be 
integrated numerically, they are not guaranteed to 
terminate. The other two algorithms are based on algebraic geometry and cannot lead to an infinite 
recursion by construction. 
While the original strategy by Kaneko and Ueda \cite{Kaneko:2009qx,Kaneko:2010kj} generated the 
lowest number of 
sectors compared to the other available strategies \cite{Hepp:1966eg,Bogner:2007mn,Bogner:2008ry,Smirnov:2008aw}, its resulting functions turned out to converge 
slower than the ones resulting from the simplest heuristic strategy. 
By contrast, our new composed strategy {\tt G2} \cite{Borowka:2015mxa} outperforms all 
others 
in terms of the number of sectors created during the decomposition, see \cite{Borowka:2015mxa}, 
and in terms of convergence during numerical integration.  

After the factorization of poles, an analytical continuation of the integrand into the complex plane is 
performed if applicable, see Ref.~\cite{Borowka:2012yc} for details. 
The last algebraic step is the subtraction of the poles and expansion in $\eps$, 
such that the coefficients of a Laurent series in $\eps$ are obtained in the form of 
parametric functions. Up to this point, kinematic invariants are left symbolic.
Explicit values are only inserted 
at the numerical stage. This has the advantage that looping over ranges of kinematic points 
is facilitated as only the numerical integration step has to be performed repeatedly. 
On the other hand, sometimes only one or two kinematic points of highly complicated integrals 
are of interest for numerical checks. In this case, including values for the kinematic 
invariants from the start simplifies the functions which need to be integrated numerically 
at the end. Explicit numerical values can be 
specified in addition to the definition of the scalar products of the external momenta in 
{\tt ScalarProductRules} in the {\it math.m} input file, e.g. 
 {\tt ScalarProductRules=\{$\dots$, s->5.1\}}. 
%
%
\section{New features at one glance}
The following list features improvements made to \sd{} with the upgrade to version 3:
\begin{itemize}
\item Two additional deterministic sector decomposition algorithms based on computational geometry 
are implemented. \vspace{-4pt}
\item Numerators can be given in terms of inverse propagators.\vspace{-4pt}
\item Loop integrals with linear propagators can be handled.\vspace{-4pt}
\item Additional $\eps$-dependent symbolic functions can be included in parametric integrals. \vspace{-4pt}
\item The restructured user input helps in building interfaces to reduction programs 
\cite{Anastasiou:2004vj,Smirnov:2008iw,vonManteuffel:2012np,Lee:2013mka} or Loopedia \cite{Papara:1}. 
\vspace{-4pt} 
\item The two numerical integrators \textsc{CQuad} \cite{cquad} and \textsc{NIntegrate} \cite{wolfram:1} 
are included in addition to the implementation of the updated version of \textsc{Cuba} 
\cite{Hahn:2004fe,Agrawal:2011tm,Hahn:2014fua}. \textsc{CQuad} 
is the fastest adaptive one-dimensional parameter integrator on the market and is chosen automatically 
by \sd{} when it can be used.  \vspace{-4pt} 
\item The usage of batch systems is facilitated and scans over parameter ranges are accelerated. \vspace{-4pt}%
\end{itemize}
\section{Summary}
These proceedings highlight the applicability of the publicly available program \sd. The operational 
sequence  of the latter and its new features are summarized and suggestions for an optimal usage of 
the program are given. 
%
\subsection*{Acknowledgements}
The author would like to thank Gudrun Heinrich, Stephen Jones, \\
Matthias Kerner, Johannes Schlenk 
and Tom Zirke for the fruitful collaboration, and the organizers of ``Matter to the deepest 2015'' for 
their excellent work in setting up an interesting conference. 
This work is supported by the ERC Advanced Grant MC@NNLO (340983).

\bibliography{borowka_bibs}

\end{document}